\documentclass{emulateapj}

\usepackage{aas_macros}
\usepackage{natbib}

\usepackage{epsfig}  

\newcommand{\mlapm}{\texttt{MLAPM}}

\newcommand{\Tab}[1]{Table~\ref{#1}}
\newcommand{\Sec}[1]{Section~\ref{#1}}
\newcommand{\Eq}[1]{Eq.~(\ref{#1})}
\newcommand{\Fig}[1]{Figure~\ref{#1}}
\newcommand{\hMpc}{{\ifmmode{h^{-1}{\rm Mpc}}\else{$h^{-1}$Mpc}\fi}}
\newcommand{\hkpc}{{\ifmmode{h^{-1}{\rm kpc}}\else{$h^{-1}$kpc}\fi}}
\newcommand{\hMsun}{{\ifmmode{h^{-1}{\rm {M_{\odot}}}}\else{$h^{-1}{\rm{M_{\odot}}}$}\fi}}

\def\vecnabla{{\nabla}}
\def\lesssim{\mathrel{\hbox{\rlap{\hbox{\lower4pt\hbox{$\sim$}}}\hbox{$<$}}}}
\def\gtrsim{\mathrel{\hbox{\rlap{\hbox{\lower4pt\hbox{$\sim$}}}\hbox{$>$}}}}
\def\LCDM{$\Lambda$CDM}

\def\beq{\begin{equation}}
\def\eeq{\end{equation}}
\def\bey{\begin{eqnarray}}
\def\eey{\end{eqnarray}}

\def\Msun{M_\odot}

\def\a0{$a_0$}

\begin{document}

\title[Phantom Dark Matter]
      {On the separation between baryonic and dark matter: evidence for phantom dark matter?}
\author
{Alexander Knebe\altaffilmark{1}, 
Claudio Llinares\altaffilmark{2},
Xufen Wu\altaffilmark{3}
HongSheng Zhao\altaffilmark{3,4}}

\altaffiltext{1}{Departamento de Fisica Teorica, Modulo C-XI, Facultad de Ciencias, Universidad Autonoma de Madrid, 28049 Cantoblanco, Madrid, Spain}
\altaffiltext{2}{Astrophysikalisches Institut Potsdam, An der Sternwarte 16, 14482 Postdam, Germany}
\altaffiltext{3}{SUPA, University of St Andrews, North Haugh, Fife, KY16 9SS, UK}
\altaffiltext{4}{Leiden University, Sterrewacht and Instituut-Lorentz, Niels-Bohrweg 2, 2333 CA, Leiden, The Netherlands}

\renewcommand{\topfraction}{0.85}
\renewcommand{\textfraction}{0.1}

\date{Submitted Version ...}



\label{firstpage}

\begin{abstract}
  The recent years have seen combined measurements of X-ray and (weak)
  lensing contours for colliding galaxy clusters such as, for
  instance, the famous ``Bullet'' cluster. These observations have
  revealed offsets in the peaks of the baryonic and (dominant)
  gravitational matter component of order $\sim100-200$ kpc. Such
  discrepancies are difficult to explain using modified theories for
  gravity other than dark matter. Or are they not? Here we use
  the concept of ``phantom dark matter'' that is based upon a
  Newtonian interpretation of the MONDian gravitational potential. 
  We show that this idea is in fact capable of producing
  substantial offsets in idealistic density configurations, involving a uniform external field. 
  However, when analysed in a MONDian cosmological framework
  we  deduce that the size (and probablity) of the effect is too small to
  explain the observed offsets found in the most recent observations,
  at least in the simplest incarnation of phantom dark matter as
  discussed here.  The lensing centers in merging galaxy clusters are
  likely very close to the centers of true mass even in a MONDian
  cosmology.  This gives the support to the idea that neutrino-like
  non-collisional matter might be responsible for the observed offsets
  of lensing and X-ray peaks.
   
\end{abstract}


\section{Introduction}
\label{sec:introduction}

The standard $\Lambda$ Cold Dark Matter ($\Lambda$CDM) model \citep[cf.,][]{Komatsu08} explains the
formation of cosmological structure in the non-linear regime in a
hierarchical way, i.e. big structures are not formed monolithically
but by successive merging of small structures
\citep[e.g,.][]{Davis85}. Recent cosmological simulations also support
this idea of hierarchical structure formation in MOND gravity
\citep{Llinares08} (but see also the analytical models of
\citet{Sanders08} and \citet{Zhao08b}).  The hierarchical merging
scenario naturally promotes the picture that we should observe
collisions of (clusters of) galaxies. Observationally there is
evidence that some of these impacts actually occur with speeds that
are not readily reproduced by simulations of $\Lambda$CDM structure
formation, in the sense that the relative speed of the merging dark
halos is rarely very higher than the internal dispersion of each halo
\citep{Hayashi06,Knebe08wdm, Llinares08, Angus08}. There is, for instance, the
famous ``Bullet cluster'', an extremely high velocity merger between
two galaxy clusters, with an inferred shock velocity of
$\sim4700$km/sec. While relative encounters with comparable velocities
are rather rare in pure dark matter simulations
\citep[e.g.][]{Hayashi06, Knebe08wdm}, they may nevertheless be
accomodated when considering explicit hydrodynamical modelling of the
phenomenon \citep{Springel07}.

One unavoidbale consequence of any high speed collision of mass
concentrations seems to be the decoupling or offsetting of the
baryonic component from the dark component.  Aside from the
aforementioned Bullet cluster -- whose offset has been measured to be
approximately $\sim100$ kpc \citep[e.g. ][]{Clowe06} -- more examples
are given in \citet{Jee05a}, \citet{Jee05b}, and \citet{Bradac08}. The
latter authors actually present data for a particulr cluster
(i.e. MACS J0025.4-1222) with an even greater difference of
$\sim200$~kpc in the peak of the baryonic and the dark matter. All
this work is culminated in one particular recent \textit{Letter} by
\citet{Shan09} where a sample of 38 galaxy clusters has been studied
utilizing both X-ray and strong lensing observations. They show that
such offsets are a common phenomenon in galaxy clusters: they found at
least 13 objects with a separation greater than 50~kpc with 3 clusters
exhibiting a separation between baryonic and (hypothetic) dark
component in excess of 200~kpc.

All these papers are analysing combined X-ray and lensing observations to decipher (and actually measure) the offset between baryonic and dark matter. But how certain are we that the lensing signal is caused by ``real'' dark matter particles? What if the gravitational potential is not generated by Newtonian physics yet interpreted in that way?

One theory capable of producing potentials akin to dark matter particles is modified Newtonian dynamics \citep[MOND, ][]{Milgrom83}; (pedagogical) reviews of the concepts and successes can be found in, for instance, \citet{Sanders02} and \citet{Milgrom08}.  While MOND was originally proposed as an alternative to Newtonian gravity designed to solely explain galactic dynamics without the need for dark matter the theory has gained substantial momentum during the past decade: although current cosmological observations point to the existence of vast amounts of non-baryonic dark matter in the Universe \citep[e.g.][]{Komatsu08}, not all of the features of CDM models appear to match observational data (e.g., the ``missing satellite problem'' \citep{Klypin99, Moore99} and the so-called ``cusp-core crisis'' \citep[e.g.][]{deBlok03,Swaters03}). Just as for CDM the MOND theory successfully matches observations on a wide range of scales, different types of galaxies including dwarfs and giants, spirals and ellipticals \citep{Famaey05,Gentile07b,Milgrom07a,Milgrom07b,Sanders07,McGaugh08,Angus08b}. However, one of MOND's major set-backs for a long time was the lack of a covariant formulation of the theory. This has been remedied by \citet{Bekenstein04} who was the first to cast MOND into a more universal form compliant with general relativity. This in turn spawned further investigations into the same direction leaving us nowadays with various relativistic formulations of the MOND theory \citep[e.g.][]{Bekenstein04,Sanders05,Bruneton07,Zhao07,Zlosnik07,Zhao08,Skordis08,Blanchet09}; a recent review of both MOND and its relativistic offspring (in particular the \textit{TeVeS} formulation of \citet{Bekenstein04}) can be found in \citet{Milgrom08} and \citet{Skordis09}. We though need to acknowledge that despite the original idea of abandoning the need for dark matter, even MOND cannot do without it: despite the great success we need to accept that even MOND cannot do well without dark matter completely. A recent study utilizing a combination of strong and weak lensing by galaxy clusters indicates the necessity for neutrinos of mass $5-7eV$ \citep{Natarajan08}. And to be consistent with dark matter estimates of galaxy clusters and observations of the CMB anisotropies \citet{Angus09a,Angus09b} claims for $11eV$ neutrinos. One theory capabale of accomodating both these requirements is that of a mass-varying neutrino by \citet{Zhao08}. In summary, we are eventually left with a situation where the development of several frameworks for a relativistic formulation of MOND enabled the study of the cosmic microwave background \citep{Skordis05,Li08mond}, cosmological structure formation \citep{Halle08,Skordis08}, strong gravitational lensing of galaxies \citep{Zhao06, Chen06,Shan08}, and weak lensing of clusters \citep{Angus07,Famaey08}. The MOND theory has matured and became a credible competitor to the commonly accepted CDM model.

In that regards, however, it appears important to look for even more tests that are capable of discriminating between MOND and Newtonian gravity, especially in the context of cosmology. We therefore raise the question whether the kinds of offsets alluded to above can be explained by simply interpreting the MONDian potential in a Newtonian way? This concept of ``phantom dark matter'' was originally introduced in the beginnings of MOND already by \citet{Milgrom86} and has recently been discussed by \citet{Milgrom08}, \citet{Wu08}, and \citet{Bienayme09}. It is based upon the idea to use the MONDian potential in a dark matter context, i.e. given the MONDian potential one can use the Newtonian Poisson's equation to derive the corresponding density of matter that would be needed in the Newtonian context. Then, subtracting the visible (baryonic) matter one obtains the ``virtual'' dark matter or, in other words, ``phantom dark matter'' distribution predicted by MOND. And in a Newtonian interpretation this phantom dark matter would be responsible for the gravitational lensing signal alluded to above.

We though need to acknowledge that \citet{Brownstein07} already
pointed out the possibility that the observed offset in the (alleged)
dark matter and baryonic density peaks of the Bullet cluster system can
be explained by extending the equations for gravitational lensing
to modified gravity, without the need for a dominant dark matter
component. Further, as MOND is a non-linear theory it is not clear whether the (baryonic)
matter will be distributed ab initio in the same way as phantom dark matter. We
therefore set out to answer the question whether or not these two
density fields share peaks at the same locations and are distributed
in comparable ways, respectively. Can an offset between the dark and
baryonic matter be explained by the non-linearity of the MONDian
Poisson's equation and the existence of this putative phantom matter,
respectively? 

We need to close with a cautionary note: this work does
\textit{not} deal with (collisions of) galaxies or galaxy clusters; we
are solely focusing on the properties of the matter density fields and
their respective peaks. The primary question we set out to answer is
whether or not MOND will produce offsets between the actual (baryonic)
matter component and the phantom matter field, even
though this work is motivated by observations of such offsets in
galaxy clusters.

\section{The Non-Cosmological Framework}
\label{sec:noncosmology}
Before investigating phantom dark matter in a cosmological environment we start off with phrasing the question about shifts in the respective density peaks for MONDian systems in a non-cosmological context. This will provide us with a gauge whether or not we should actually expect to find the reputed offsets.

\subsection{Phantom Dark Matter}
The MONDian Poisson's equation embedded within an external field reads as follows

\beq \label{efpoisson}
-\nabla \cdot \left[ \mu \left({|\textbf{g}|\over a_0}\right) {\bf g} \right]=4\pi G\rho,\qquad {\bf g}={\bf g}_{\rm ext} - {\bf \nabla} \Phi_{\rm int} ,
\eeq
where $\rho$ is the baryonic matter, ${\bf g}_{\rm ext}$ an external field, and $\Phi_{\rm int}$ the (internal) potential of the system and $\mu(x)$ is $\mu\rightarrow1$ for $x\gg 1$ (Newtonian limit) and
$\mu\rightarrow x$ for $x\ll 1$ (deep MOND limit).\footnote{Please note that we used $\mu(x) = x (1+x^2)^{-1/2}$ as originally suggested by \cite{Milgrom83} throughout our tests.}. We now take the
liberty to interpret this internal potential $\Phi_{\rm int}$ within the context of Newtonian gravity
\beq \label{pdm}
\nabla^2 \Phi_{\rm int}=4\pi G(\rho+\rho_{\rm ph}),
\eeq
where the $(\rho+\rho_{\rm ph})$ is the total dynamical mass of the system, and $\rho_{\rm ph}$ is the so called ``phantom dark matter'', which can be held responsible for the ``extra gravity beyond the baryonic matter''\footnote{Or in other words ``dark matter''.} in the linear Poisson's equation \Eq{pdm}. But as opposed to the dark matter theory, MOND immediately predicts the distribution of the dynamical mass as soon as the baryons $\rho$ and the gravity of the environment ${\bf g}_{\rm ext}$ are specified. However, due to the external field ${\bf g}_{\rm ext}$ the boundary condition of the (internal) system changes, not necessarily preserving spherical symmetry: the distribution of the dynamical mass is somewhat different from that of CDM halos \citep[see][]{Wu07, Wu08}. Further, there exist negative solutions of the phantom dark matter, and the peaks of dynamical mass can in fact be offset to the baryonic peaks! These effects are most significant at places where the external and internal fields are comparable and should be quantified more carefully in the following subsection.

\subsection{The Simulations}
For the non-cosmological settings studied in this Section we use the MONDian Poisson solver developed by the Bologna group \citep{Ciotti06, Nipoti07} to solve \Eq{efpoisson} and hence derive the internal potential $\Phi_{\rm int}$ of the systems under investigation. The Poisson solver is a spherical grid code, and our choice for the grid parameters is $n_r \times n_\theta \times n_\phi = 256 \times 64 \times 128$ with a radial grid spacing given by $r_i=r_0 \tan \left[(i+0.5){0.5\pi /(n_r+1)}\right]$~kpc. We further utilize
\Eq{pdm} to derive the dynamical mass and the phantom matter, respectively. 

All our (baryonic) galaxies have a Plummer density profile
\beq
\rho(r)=\left({3M\over 4\pi b^3}\right)\left(1+{r^2\over b^2}\right)^{-5/2},
\eeq
with a core radius b of 1.0~kpc. \\

We investigated several scenaria (cf. discussion below in \Sec{sec:peak_offsets}), however, decided to only present the results for one representative configuration where a single galaxy G is embedded within a strong constant external field EF. A summary of the actual parameters of the model can be found Table~\ref{models}.

\begin{table}
\makeatletter\def\@captype{table}\makeatother
\caption{Parameters of our non-cosmological test model. The mass is in unit of $(10^{10}\Msun)$, positions and $r_0$ are in units of $kpc$, and the external gravity acceleration is in unit of $a_0$.}
\begin{center}
\begin{tabular}{lclcccccc}
\hline
Model & Mass &Centre & $g_{\rm ext}$ & $r_0$ \\
\hline
G+EF   & 10.0 &  (0,0,0) & 1.0 & 2.0 \\
\hline
\\
\end{tabular}\label{models}
\end{center}
\end{table}

\subsection{Density Peak Offsets}
\label{sec:peak_offsets}
\begin{figure}{}
\makeatletter\def\@captype{figure}\makeatother
\resizebox{8cm}{!}{\includegraphics{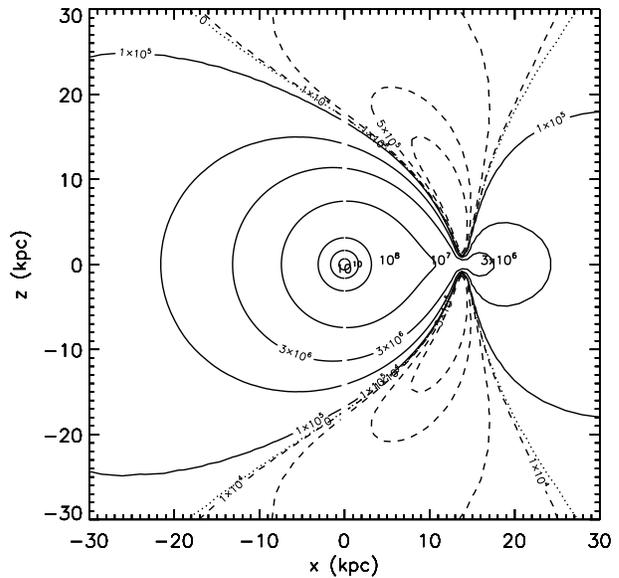}}
\caption{Isodensity countours of the dynamical mass, i.e. baryons + phantom dark matter density, on the $x-z$ plane for a galaxy embedded within an external field along the $x$-axis.}\label{efeqden1}
\end{figure}

\Fig{efeqden1} shows the distribution of the dynamical massfor model G+EF, i.e. a single galaxy embedded within an external field of strength $|{\bf g}_{\rm ext}|=a_0$ along the $x$-axis. We notice that when the internal and external field are of the same order of magnitude, i.e. at $x\approx$~15~kpc there are some noticable effects: first, we obtain negative dynamical mass on the corns perpendicular to the direction of external field also mentioned by \citet{Milgrom86} and \citet{Wu08}; second, there exist an additional peak on the $x$-axis, right where the external and internal fields cancel each other! However, this additional peak is four orders of magnitudes smaller than the baryonic peak. 

We have also run idealistic simulations with (un-)equal mass galaxies with and without external fields (not shown though).  In summary, we have seen that in most of the situations the interpretation of the MONDian potential in a Newtonian sense will lead to the prediction of additional peaks in the distribution of the dynamical mass when compared to the actual (underlying) baryonic matter distribution. However, the strength of these extra peaks vary and depend on the actual setup of the system ranging from as low as four orders of magnitude smaller to as large as 1\% for the cases considered here.  

Encouraged by the observation that we actually recover offsets in our controlled experiments we may now rightfully ask the question {\it whether these additional phantom peaks occur in realistic cosmological simulations}, i.e., whether a self-consistent cosmological simulation will provide a suitable variety of configurations so that we will in fact be able to observe (and quantify) the offset between the baryonic and phantom matter peaks.

\section{The Cosmological Framework}
\label{sec:cosmology}

\subsection{Phantom Dark Matter}
\label{sec:phantomdarkmatter}
The equation for the (MONDian) gravitational potential in a cosmological setting is somewhat different to \Eq{efpoisson} and reads

\begin{equation} \label{eq:PoissonMOND} 
\nabla\cdot\left[
  \mu\left(\frac{|\nabla\Phi_{M}|}{a \gamma(a)}\right)\nabla\Phi_{M}\right]=\frac{4\pi G}{a}
\left(\rho-\bar\rho\right) \ , 
\end{equation}

\noindent
We further took the
liberty to encode the MONDian acceleration scale $\gamma(a)$ as a
(possible) function of the cosmic expansion factor $a$. The most naive
choice would be $\gamma(a) = g_0 = 1.2\times 10^{-8}{\rm cm}/{\rm
  sec}^2$ whereas other theories may lead to different dependencies;
for instance, in \citet{Zhao08} $\gamma(a)$ is given as
$\gamma(a)=a^{1/2}g_0$. For more details and a derivation of this
equation we refer the reader to \citet{Llinares08} where it has been justified and
implemented into the cosmological $N$-body code \texttt{MLAPM} \citep{Knebe01}.

Given the MONDian potential $\Phi_M$ we may now apply the same logic as in \Sec{sec:noncosmology} and use it
with the Newtonian Poisson's equation whose
right-hand-side will again no longer be the (baryonic) density $\rho$ field
alone but rather read as follows

\begin{equation}\label{eq:phantomdensity}
  \displaystyle \vecnabla \cdot \left[ \nabla\Phi_M \right]
  = \frac{4 \pi G}{a} \left[ (\rho+\rho_{\rm ph}) - \overline{(\rho+\rho_{\rm ph})} \right] \ .
\end{equation}

\noindent
This is the defining equation for the phantom matter density
field $\rho_{\rm ph}$ used througout this Section.

\subsection{The Simulation}
\label{sec:simulation}
The analysis presented in here is based upon a
particular simulation published in \citet{Llinares08}, i.e. the
OCBMond2 model. This simulation has been run in a cosmological volume
with a side length of $32h^{-1}$Mpc and utilized $128^3$ particles. It
employed the MONDification of the $N$-body code \texttt{MLAPM}
\citep{Knebe01, Llinares08}. We chose to simulate an open universe
with neither dark matter nor dark energy but characterized by
$\Omega_{b}=0.04$. The simulation was started at redshift $z=50$ and
resorted to a Hubble parameter of $h=0.7$. We further need to mention
that there are two values $\sigma_8$ in a MOND simulation, one
characterising the amplitude of fluctuations of the initial condition
and one measuring the strength of fluctuations at the present
time. This comes about because of the faster growth of structures in
MOND \citep[cf.][]{Sanders01, Knebe04b, Llinares08}, i.e. in order to
arrive at a comparable evolutionary stage to a $\Lambda$CDM model at
redshift $z=0$ with $\sigma_8 \sim 0.9$ we had to lower the magnitude
of the fluctuations to $\sigma_8=0.4$ during the process of generating the initial condition. We
acknowledge that this value is incompatible with CMB constraints, at
least in the dark matter explanation for cosmic structure
formation. MOND, however, is a highly non-linear theory and the
simulation presented and used here should be considered as a first toy
model for trying to understand structure formation using modified
gravity. For more details and elaborate study of the simulation we refer
the reader again to \citet{Llinares08}.

In order to perform the analysis presented here we further modified our potential solver to use the MONDian potential $\Phi_M$
obtained by solving \Eq{eq:PoissonMOND} for the final output at today's redshift $z=0$ and our knowledge about the (baryonic) matter density $\rho$ together with \Eq{eq:phantomdensity} to obtain the resulting phantom density $\rho_{\rm ph}$.

\subsection{Locating Density Peaks}
\label{sec:peaks}
\begin{figure}
\begin{center}
\begin{minipage}{0.49\textwidth}
        \epsfig{file=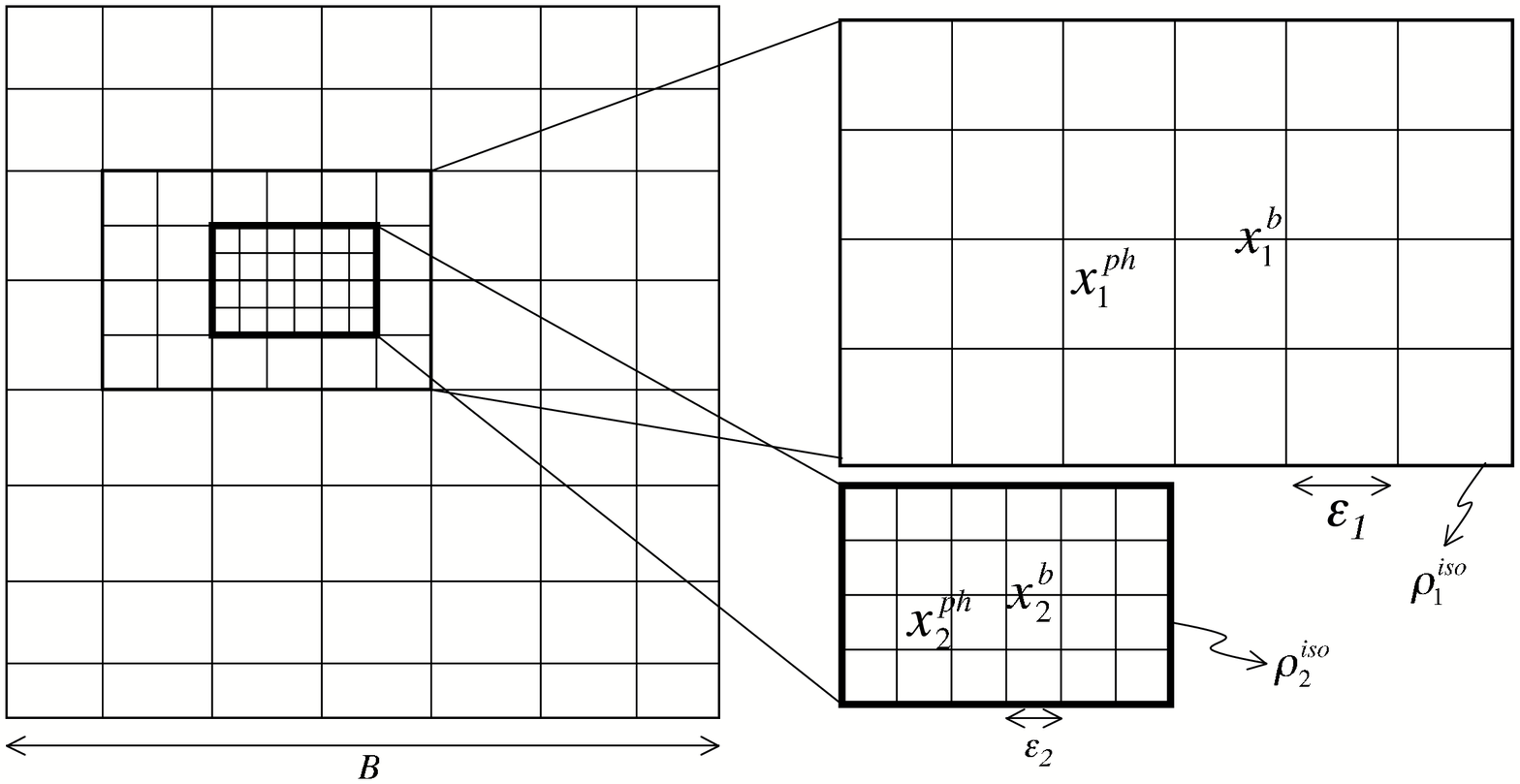, width=0.9\textwidth, angle=0}
\end{minipage}
\end{center}
\caption{Sketch illustrating the definition of the (baryonic) matter
  density centres $x^b_i$ on various refinement patches and the
  corresponding phantom matter peaks $x^{\rm ph}_i$.  Note that the
  boundary of each patch is an isodensity contour $\rho^{iso}_i$ in
  the (baryonic) matter distribution. Due to the nature of the mass
  assignment of the (baryonic) particles onto each refinement patch we
  are left with a density field smoothed on approximately the scale of
  the respective grid spacing $\epsilon_i$.\label{fig:PhDM}}
\end{figure}

Given both the (baryonic) matter density $\rho$ and the phantom matter
density $\rho_{\rm ph}$ we determine peaks in both fields. To this
extend we smooth the fields on various scales and study spatial
offsets in corresponding peaks in relation to the smoothing scale.
This is accomplished by exploiting the adaptive mesh nature of the
simulation code \mlapm\ used to generate the simulation in the first
place \citep{Knebe01, Llinares08}: the (baryonic) matter is
represented by discrete particles whose mass is assigned to a regular
grid covering the whole computational volume. This grid is then
recursively refined in regions of high (baryonic) density according to
a pre-selected refinement criterion of 4, 8, or 16 particles per
cell. This leaves us with a hierarchy of (nested) refinement patches
where the boundary of each such patch defines a unique (baryonic)
isodensity contour. The situation is illustrated in \Fig{fig:PhDM}
where we show an example of two nested refinements embedded within a
regular domain grid covering the whole computational volume of side
length $B$. For each isolated patch we calculate $x^b_i$ and $x^{\rm
  ph}_i$ by first finding the position of the cell containing the
maximum in $\rho$ ($\rho_{\rm ph}$); we then use the density (phantom
density) weighted average of the 27 neighbouring cells to define
$x^b_i$ ($x^{\rm ph}_i$). The relevant quantity to be studied below is
the difference between the two peaks in the density field

\begin{equation} \label{eq:offset}
 D = | x^b_i - x^{\rm ph}_i | \ ,
\end{equation}

\noindent
that obviously is both a function of the smoothing scale $\epsilon_i$
and the isodensity contour $\rho^{\rm iso}_i$.

\subsection{Density Peak Offsets}
\label{sec:offsets}
\begin{table}
\begin{center}
  \caption{Smoothing scales in \hkpc.}
\begin{tabular}{rr}
\hline
$L$     & $\epsilon_{\rm L}$ [\hkpc] \\
\hline
16382   & 1.95\\\
8192    & 3.91\\
4096    & 7.81\\
2048    & 15.63\\
1024    & 31.25\\
512     & 62.50\\
\end{tabular}
\label{tab:smoothingscales}
\end{center}
\end{table}

\begin{figure}
\begin{center}
\begin{minipage}{0.47\textwidth}
        \epsfig{file=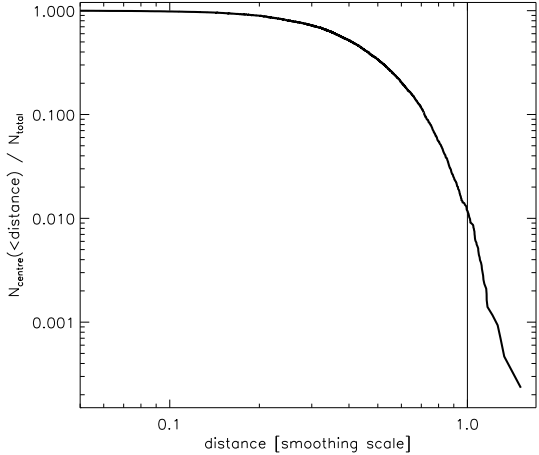, width=0.9\textwidth, angle=0}
\end{minipage}
\end{center}
\caption{The cumulative number distribution $N(<D)$ of the offset $D$
  between (baryonic) matter density and phantom density. The offset
  has been normalized to the respective smoothing scale of the
  refinement patch it is based upon. \label{fig:NdistNorm}}
\end{figure}

We are primarily interested in the question whether or not there are
any (substantial) offsets in the peaks of the (baryonic) matter
density field and the corresponding phantom field defined via
\Eq{eq:phantomdensity}. For the time being we therefore ignore any
relation this offset has with the refinement level it is based upon
(cf. \Eq{eq:offset}). In \Fig{fig:NdistNorm} we simply plot the
cumulative distance distribution $N(<D)$ normalized to the total
number of refinement patches on all levels; we further chose to
normalize the distance to the respective smoothing scale $\epsilon_i$
as we consider distances smaller than this scale to be below the
resolution limit and hence not credible. The physical values for
$\epsilon_i$ for the grid levels used in our calculations are
summarized in \Tab{tab:smoothingscales}.

While most of the offsets between (baryonic) matter and phantom matter
are in fact smaller than the resolution limit there are nevertheless
of order 1\% instances for which we observe larger (and hence
physical) differences!

We like to caution the reader that the same matter peak enters
multiple times (at most six times) into \Fig{fig:NdistNorm} (and all
subsequent plots below). This is due to the fact that we smooth the
same peak using various smoothing scales $\epsilon_i$ listed in
\Tab{tab:smoothingscales}. However, as we are not interested in the
change in offset for a given peak when altering the smoothing scale we
can treat them independently.

\begin{figure}
\begin{center}
\begin{minipage}{0.47\textwidth}
        \epsfig{file=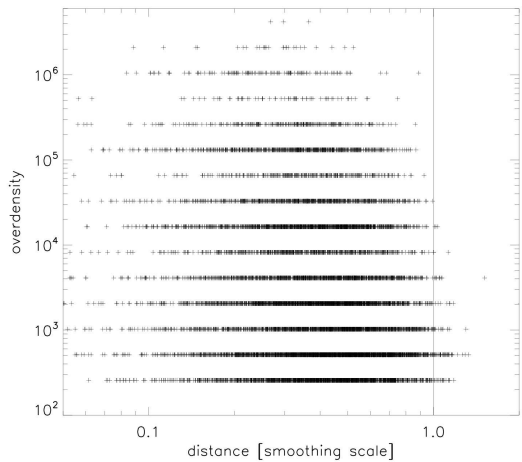, width=0.9\textwidth, angle=0}
\end{minipage}
\end{center}
\caption{The relation between the (baryonic) isodensity level of the
  refinement patch and the (normalized) offset between matter and
  phantom density on it. \label{fig:DistOvdensNorm}}
\end{figure}

Even though we just found that there is a small yet measurable
probability of finding an offset between baryonic and phantom density
it still remains unclear how this can be interpreted in terms of
astrophysical objects. Observationally the edge of an object is
primarily defined by a given threshold in (over-)density. This,
however, is a natural by-product of our method for calculating $D$
(cf. \Sec{sec:offsets}). We therefore plot in \Fig{fig:DistOvdensNorm}
the dependence of the (normalized) distance $D$ on the overdensity of
the corresponding refinement patch. Recall that the usual overdensity
limit for virialized objects in an \LCDM\ cosmology at redshift $z=0$
is $\approx340$ and approximately coincides with our coarsest
refinement level.

As the only credible difference in the position of (baryonic) matter
and phantom matter peaks are apparent on the lower iso-density levels
(i.e. the physically larger refinement patches) one may raise the
question whether we calculated the offset for corresponding peaks. A
large refinement patch will certainly host several peaks both in
baryonic and phantom matter, so how can we be sure to take the
difference between matching peaks? Maybe the maximum baryonic density
is not at the same position as the maximum phantom density
(cf. \Sec{sec:offsets})? This concern is readily eliminated as the
maximum offset observed is no larger than two times the actual
smoothing scale, i.e. the peaks lie in two neighbouring grid cells.

\begin{figure}
\begin{center}
\begin{minipage}{0.47\textwidth}
        \epsfig{file=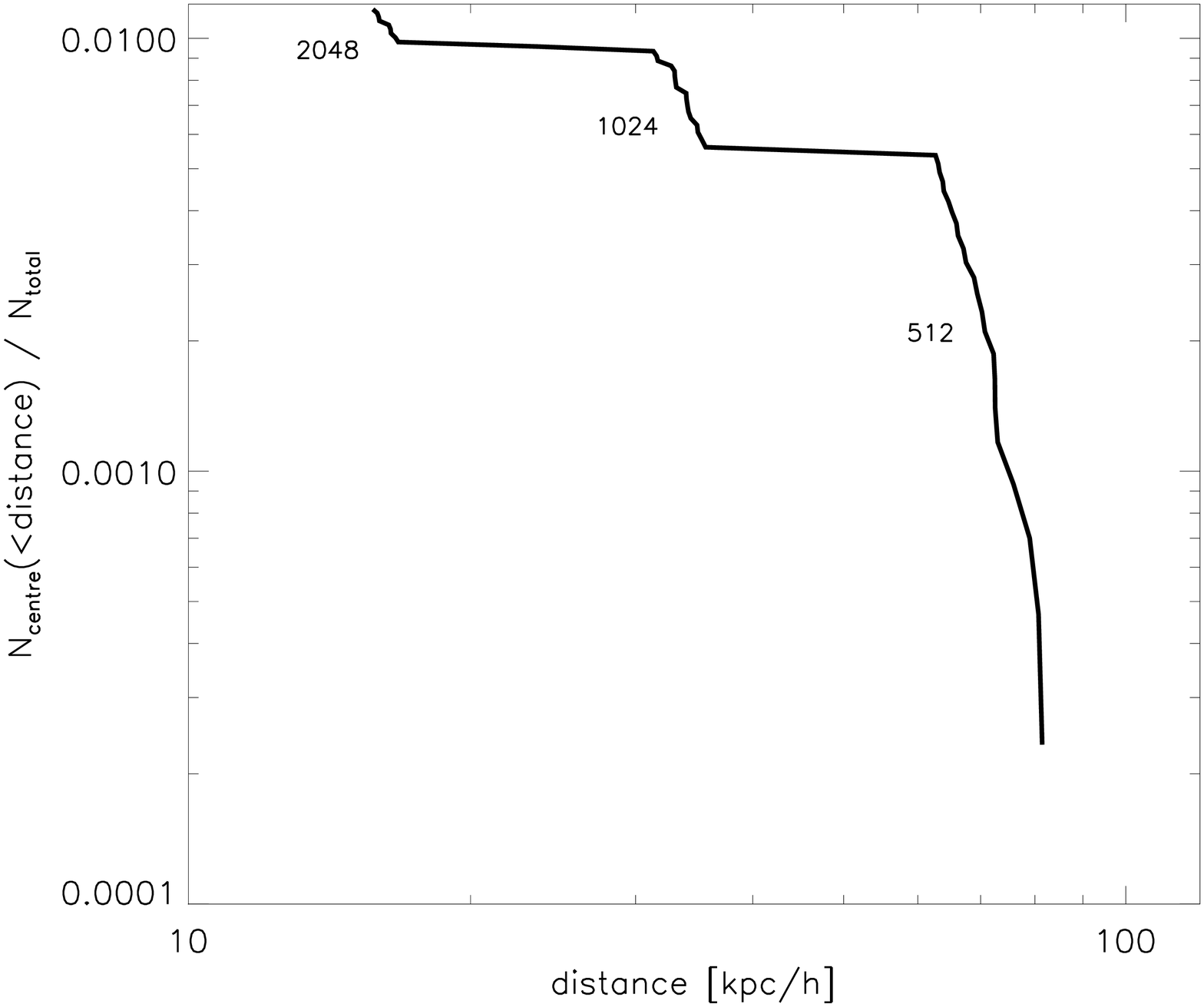, width=0.9\textwidth, angle=0}
\end{minipage}
\end{center}
\caption{Cumulative distribution of isolated patches with an offset
  $D>\epsilon_L$ (normalized to the total number of patches). Only
  patches on grids $L\leq2048$ fulfill this criterion and the
  contribution of these grids to the distribution are marked by the
  respective $L$ values given in the plot. \label{fig:Ndist-kpc}}
\end{figure}

So far, we always normalized the offset $D$ to the respective
smoothing scale $\epsilon_i$. However, to gain a better and more
quantitative feeling for the relevance of our results it appears
obligatory to also consider the distance in physical units \hkpc. To
this extent we plot in \Fig{fig:Ndist-kpc} the cumulative distribution
of all offsets $D>\epsilon_L$ larger than the respective smoothing
scale in physical \hkpc.\footnote{Note that this figure represents a
  zoom of \Fig{fig:NdistNorm} into the region right of the vertical
  line and in physical units on the $x$-axis now.} Note that only the
three coarsest grids (i.e. $512^3$, $1024^3$, and $2048^3$) lead to
offsets that are larger the smoothing scale; for all finer grids the
distance in the (baryonic) matter and the phantom matter peak is below
the credibility level given by the smoothing scale. We further observe
that the reasonable offsets lie in the range between 15--80\hkpc. We
though need to acknowledge that the absolute fraction of isolated
patches fulfilling this credibility criterion is below 0.6\%. Our
simulation therefore has difficulties to accomodate offsets as large
as the ones observed.

\section{Summary and Conclusions}
\label{sec:summary}
Driven by the observation of offsets in the baryonic and gravitational
matter distribution in collisions of galaxy clusters
\citep[e.g. ][]{Shan09} we explore such phenomena in the context of
phantom dark matter. This is an interpretation of the MONDian
potential (generated purely by baryons) in a Newtonian context,
i.e. the MOND potential is used with the standard (Newtonian) Poisson's equation
and the resulting right-hand-side source term is understood as a
combination of the baryonic matter and some phantom dark matter.

An initial study of (interacting) galaxies in isolation as well as external fields indicated that
we should expect to find additional peaks in the distribution of the dynamical mass
as opposed to the baryonic mass distriution. However, the strength of these extra peaks
varied and the contrast may in some cases be even too low to be observed.

Utilizing a MONDification of the $N$-body code \texttt{MLAPM}
\citep{Knebe01} we set ourselves into the position of calculating both
the baryonic and phantom density distributions in a fully self-consistent MONDian
cosmological simulation on adaptive refinement
patches. We then quantified differences in the peaks of both these
fields concluding that the (theoretically predicted) offsets are too
small to be compliant with the observed offsets, at least in the
presented incarnation of phantom matter and our MONDian cosmological
simulation.

One possible drawback of our applied method is the fact that the
isodensity level that define isolated refinement patches are based
upon $\rho_b$ only. However, we compensated for this quibble by
adjusting the refinement criterion and subsequentially modifying the
size of isolated patches; we though could not detect any systematics.

We conclude that our results give support to the idea that neutrino-like non-collisional matter might be responsible for the observed offsets of lensing and X-ray peaks. There are in fact indications by several authors that non-classical neutrinos are required to explain phenomena, such as, cluster lensing \citep{Natarajan08} or CMB anisotropies \citep{Angus09b} within the context of (relativistic) MOND. One theory capabale of accomodating both these requirements is that of a mass-varying neutrino by \citet{Zhao08} to be studied more detailed in future work.
 
\acknowledgments
AK is supported by the MICINN through the Ramon y Cajal programme. CL
and AK further acknowledge funding by the DFG under grant KN
755/2. This work was also carried out under the HPC-EUROPA++ project
(project number: 211437), with the support of the European Community -
Research Infrastructure Action of the FP7 ``Coordination and support
action'' Programme. XW thanks the support of the SUPA studentship.

\bibliographystyle{apj}
\bibliography{archive}

\begin{thebibliography}{58}
\expandafter\ifx\csname natexlab\endcsname\relax\def\natexlab#1{#1}\fi

\bibitem[{{Angus}(2008)}]{Angus08b}
{Angus}, G.~W. 2008, \mnras, 387, 1481

\bibitem[{{Angus}(2009)}]{Angus09a}
---. 2009, \mnras, 394, 527

\bibitem[{{Angus} {et~al.}(2009){Angus}, {Famaey}, \& {Diaferio}}]{Angus09b}
{Angus}, G.~W., {Famaey}, B., \& {Diaferio}, A. 2009, ArXiv e-prints

\bibitem[{{Angus} \& {McGaugh}(2008)}]{Angus08}
{Angus}, G.~W., \& {McGaugh}, S.~S. 2008, \mnras, 383, 417

\bibitem[{{Angus} {et~al.}(2007){Angus}, {Shan}, {Zhao}, \& {Famaey}}]{Angus07}
{Angus}, G.~W., {Shan}, H.~Y., {Zhao}, H.~S., \& {Famaey}, B. 2007, \apjl, 654,
  L13

\bibitem[{{Bekenstein}(2004)}]{Bekenstein04}
{Bekenstein}, J.~D. 2004, \prd, 70, 083509

\bibitem[{{Bienaym{\'e}} {et~al.}(2009){Bienaym{\'e}}, {Famaey}, {Wu}, {Zhao},
  \& {Aubert}}]{Bienayme09}
{Bienaym{\'e}}, O., {Famaey}, B., {Wu}, X., {Zhao}, H.~S., \& {Aubert}, D.
  2009, ArXiv e-prints

\bibitem[{{Blanchet} \& {Le Tiec}(2009)}]{Blanchet09}
{Blanchet}, L., \& {Le Tiec}, A. 2009, \prd, 80, 023524

\bibitem[{{Brada{\v c}} {et~al.}(2008){Brada{\v c}}, {Allen}, {Treu},
  {Ebeling}, {Massey}, {Morris}, {von der Linden}, \& {Applegate}}]{Bradac08}
{Brada{\v c}}, M., {Allen}, S.~W., {Treu}, T., {Ebeling}, H., {Massey}, R.,
  {Morris}, R.~G., {von der Linden}, A., \& {Applegate}, D. 2008, \apj, 687,
  959

\bibitem[{{Brownstein} \& {Moffat}(2007)}]{Brownstein07}
{Brownstein}, J.~R., \& {Moffat}, J.~W. 2007, \mnras, 382, 29

\bibitem[{{Bruneton} \& {Esposito-Far{\`e}se}(2007)}]{Bruneton07}
{Bruneton}, J.-P., \& {Esposito-Far{\`e}se}, G. 2007, \prd, 76, 124012

\bibitem[{{Chen} \& {Zhao}(2006)}]{Chen06}
{Chen}, D.-M., \& {Zhao}, H. 2006, \apjl, 650, L9

\bibitem[{{Ciotti} {et~al.}(2006){Ciotti}, {Londrillo}, \& {Nipoti}}]{Ciotti06}
{Ciotti}, L., {Londrillo}, P., \& {Nipoti}, C. 2006, \apj, 640, 741

\bibitem[{{Clowe} {et~al.}(2006){Clowe}, {Brada{\v c}}, {Gonzalez},
  {Markevitch}, {Randall}, {Jones}, \& {Zaritsky}}]{Clowe06}
{Clowe}, D., {Brada{\v c}}, M., {Gonzalez}, A.~H., {Markevitch}, M., {Randall},
  S.~W., {Jones}, C., \& {Zaritsky}, D. 2006, \apjl, 648, L109

\bibitem[{{Davis} {et~al.}(1985){Davis}, {Efstathiou}, {Frenk}, \&
  {White}}]{Davis85}
{Davis}, M., {Efstathiou}, G., {Frenk}, C.~S., \& {White}, S.~D.~M. 1985, \apj,
  292, 371

\bibitem[{{de Blok} {et~al.}(2003){de Blok}, {Bosma}, \& {McGaugh}}]{deBlok03}
{de Blok}, W.~J.~G., {Bosma}, A., \& {McGaugh}, S. 2003, \mnras, 340, 657

\bibitem[{{Famaey} {et~al.}(2008){Famaey}, {Angus}, {Gentile}, {Shan}, \&
  {Zhao}}]{Famaey08}
{Famaey}, B., {Angus}, G.~W., {Gentile}, G., {Shan}, H.~Y., \& {Zhao}, H.~S.
  2008, in Proceed. of the Sixth International Heidelberg Conference on Dark
  Matter in Astro- and Particle Physics (DARK 2007), held at the University of
  Sydney, Australia, 23-28 September 2007, Eds. H. V. Klapdor-Kleingrothaus, G.
  Lewis, World Scientific, Singapore, ISBN-13 978-981-281-434-0, ISBN-10
  981-281-434-5, pp. 393-401, 393--401

\bibitem[{{Famaey} \& {Binney}(2005)}]{Famaey05}
{Famaey}, B., \& {Binney}, J. 2005, \mnras, 363, 603

\bibitem[{{Gentile} {et~al.}(2007){Gentile}, {Famaey}, {Combes}, {Kroupa},
  {Zhao}, \& {Tiret}}]{Gentile07b}
{Gentile}, G., {Famaey}, B., {Combes}, F., {Kroupa}, P., {Zhao}, H.~S., \&
  {Tiret}, O. 2007, \aap, 472, L25

\bibitem[{{Halle} {et~al.}(2008){Halle}, {Zhao}, \& {Li}}]{Halle08}
{Halle}, A., {Zhao}, H., \& {Li}, B. 2008, \apjs, 177, 1

\bibitem[{{Hayashi} \& {White}(2006)}]{Hayashi06}
{Hayashi}, E., \& {White}, S.~D.~M. 2006, \mnras, 370, L38

\bibitem[{{Jee} {et~al.}(2005{\natexlab{a}}){Jee}, {White}, {Ben{\'{\i}}tez},
  {Ford}, {Blakeslee}, {Rosati}, {Demarco}, \& {Illingworth}}]{Jee05a}
{Jee}, M.~J., {White}, R.~L., {Ben{\'{\i}}tez}, N., {Ford}, H.~C., {Blakeslee},
  J.~P., {Rosati}, P., {Demarco}, R., \& {Illingworth}, G.~D.
  2005{\natexlab{a}}, \apj, 618, 46

\bibitem[{{Jee} {et~al.}(2005{\natexlab{b}}){Jee}, {White}, {Ford},
  {Blakeslee}, {Illingworth}, {Coe}, \& {Tran}}]{Jee05b}
{Jee}, M.~J., {White}, R.~L., {Ford}, H.~C., {Blakeslee}, J.~P., {Illingworth},
  G.~D., {Coe}, D.~A., \& {Tran}, K.-V.~H. 2005{\natexlab{b}}, \apj, 634, 813

\bibitem[{{Klypin} {et~al.}(1999){Klypin}, {Gottl{\"o}ber}, {Kravtsov}, \&
  {Khokhlov}}]{Klypin99}
{Klypin}, A., {Gottl{\"o}ber}, S., {Kravtsov}, A.~V., \& {Khokhlov}, A.~M.
  1999, \apj, 516, 530

\bibitem[{{Knebe} {et~al.}(2008){Knebe}, {Arnold}, {Power}, \&
  {Gibson}}]{Knebe08wdm}
{Knebe}, A., {Arnold}, B., {Power}, C., \& {Gibson}, B.~K. 2008, \mnras, 386,
  1029

\bibitem[{{Knebe} \& {Gibson}(2004)}]{Knebe04b}
{Knebe}, A., \& {Gibson}, B.~K. 2004, \mnras, 347, 1055

\bibitem[{{Knebe} {et~al.}(2001){Knebe}, {Green}, \& {Binney}}]{Knebe01}
{Knebe}, A., {Green}, A., \& {Binney}, J. 2001, \mnras, 325, 845

\bibitem[{{Komatsu} {et~al.}(2008){Komatsu}, {Dunkley}, {Nolta}, {Bennett},
  {Gold}, {Hinshaw}, {Jarosik}, {Larson}, {Limon}, {Page}, {Spergel},
  {Halpern}, {Hill}, {Kogut}, {Meyer}, {Tucker}, {Weiland}, {Wollack}, \&
  {Wright}}]{Komatsu08}
{Komatsu}, E., {Dunkley}, J., {Nolta}, M.~R., {Bennett}, C.~L., {Gold}, B.,
  {Hinshaw}, G., {Jarosik}, N., {Larson}, D., {Limon}, M., {Page}, L.,
  {Spergel}, D.~N., {Halpern}, M., {Hill}, R.~S., {Kogut}, A., {Meyer}, S.~S.,
  {Tucker}, G.~S., {Weiland}, J.~L., {Wollack}, E., \& {Wright}, E.~L. 2008,
  ArXiv e-prints, 803

\bibitem[{{Li} {et~al.}(2008){Li}, {Barrow}, {Mota}, \& {Zhao}}]{Li08mond}
{Li}, B., {Barrow}, J.~D., {Mota}, D.~F., \& {Zhao}, H. 2008, \prd, 78, 064021

\bibitem[{{Llinares} {et~al.}(2008){Llinares}, {Knebe}, \& {Zhao}}]{Llinares08}
{Llinares}, C., {Knebe}, A., \& {Zhao}, H. 2008, \mnras, 391, 1778

\bibitem[{{McGaugh}(2008)}]{McGaugh08}
{McGaugh}, S.~S. 2008, \apj, 683, 137

\bibitem[{{Milgrom}(1983)}]{Milgrom83}
{Milgrom}, M. 1983, \apj, 270, 365

\bibitem[{{Milgrom}(1986)}]{Milgrom86}
---. 1986, \apj, 306, 9

\bibitem[{{Milgrom}(2007)}]{Milgrom07b}
---. 2007, \apjl, 667, L45

\bibitem[{{Milgrom}(2008)}]{Milgrom08}
---. 2008, ArXiv e-prints

\bibitem[{{Milgrom} \& {Sanders}(2007)}]{Milgrom07a}
{Milgrom}, M., \& {Sanders}, R.~H. 2007, \apjl, 658, L17

\bibitem[{{Moore} {et~al.}(1999){Moore}, {Ghigna}, {Governato}, {Lake},
  {Quinn}, {Stadel}, \& {Tozzi}}]{Moore99}
{Moore}, B., {Ghigna}, S., {Governato}, F., {Lake}, G., {Quinn}, T., {Stadel},
  J., \& {Tozzi}, P. 1999, \apjl, 524, L19

\bibitem[{{Natarajan} \& {Zhao}(2008)}]{Natarajan08}
{Natarajan}, P., \& {Zhao}, H. 2008, \mnras, 389, 250

\bibitem[{{Nipoti} {et~al.}(2007){Nipoti}, {Londrillo}, \& {Ciotti}}]{Nipoti07}
{Nipoti}, C., {Londrillo}, P., \& {Ciotti}, L. 2007, \apj, 660, 256

\bibitem[{{Sanders}(2001)}]{Sanders01}
{Sanders}, R.~H. 2001, \apj, 560, 1

\bibitem[{{Sanders}(2005)}]{Sanders05}
---. 2005, \mnras, 363, 459

\bibitem[{{Sanders}(2008)}]{Sanders08}
---. 2008, \mnras, 463

\bibitem[{{Sanders} \& {McGaugh}(2002)}]{Sanders02}
{Sanders}, R.~H., \& {McGaugh}, S.~S. 2002, \araa, 40, 263

\bibitem[{{Sanders} \& {Noordermeer}(2007)}]{Sanders07}
{Sanders}, R.~H., \& {Noordermeer}, E. 2007, \mnras, 379, 702

\bibitem[{{Shan} {et~al.}(2008){Shan}, {Feix}, {Famaey}, \& {Zhao}}]{Shan08}
{Shan}, H.~Y., {Feix}, M., {Famaey}, B., \& {Zhao}, H. 2008, \mnras, 387, 1303

\bibitem[{{Shan} {et~al.}(2009){Shan}, {Qin}, {Fort}, {Tao}, \& {Wu}}]{Shan09}
{Shan}, H.-Y., {Qin}, B., {Fort}, B., {Tao}, C., \& {Wu}, X.-P. 2009, \apjl,
  submitted

\bibitem[{{Skordis}(2008)}]{Skordis08}
{Skordis}, C. 2008, \prd, 77, 123502

\bibitem[{{Skordis}(2009)}]{Skordis09}
---. 2009, Classical and Quantum Gravity, 26, 143001

\bibitem[{{Skordis} {et~al.}(2006){Skordis}, {Mota}, {Ferreira}, \&
  {B{\oe}hm}}]{Skordis05}
{Skordis}, C., {Mota}, D.~F., {Ferreira}, P.~G., \& {B{\oe}hm}, C. 2006,
  Physical Review Letters, 96, 011301

\bibitem[{{Springel} \& {Farrar}(2007)}]{Springel07}
{Springel}, V., \& {Farrar}, G.~R. 2007, \mnras, 380, 911

\bibitem[{{Swaters} {et~al.}(2003){Swaters}, {Madore}, {van den Bosch}, \&
  {Balcells}}]{Swaters03}
{Swaters}, R.~A., {Madore}, B.~F., {van den Bosch}, F.~C., \& {Balcells}, M.
  2003, \apj, 583, 732

\bibitem[{{Wu} {et~al.}(2008){Wu}, {Famaey}, {Gentile}, {Perets}, \&
  {Zhao}}]{Wu08}
{Wu}, X., {Famaey}, B., {Gentile}, G., {Perets}, H., \& {Zhao}, H. 2008,
  \mnras, 386, 2199

\bibitem[{{Wu} {et~al.}(2007){Wu}, {Zhao}, {Famaey}, {Gentile}, {Tiret},
  {Combes}, {Angus}, \& {Robin}}]{Wu07}
{Wu}, X., {Zhao}, H., {Famaey}, B., {Gentile}, G., {Tiret}, O., {Combes}, F.,
  {Angus}, G.~W., \& {Robin}, A.~C. 2007, \apjl, 665, L101

\bibitem[{{Zhao}(2006)}]{Zhao06}
{Zhao}, H. 2006, ArXiv Astrophysics e-prints

\bibitem[{{Zhao}(2007)}]{Zhao07}
---. 2007, \apjl, 671, L1

\bibitem[{{Zhao}(2008)}]{Zhao08}
---. 2008, ArXiv e-prints, 805

\bibitem[{{Zhao} {et~al.}(2008){Zhao}, {Xu}, \& {Dobbs}}]{Zhao08b}
{Zhao}, H., {Xu}, B.-X., \& {Dobbs}, C. 2008, \apj, 686, 1019

\bibitem[{{Zlosnik} {et~al.}(2007){Zlosnik}, {Ferreira}, \&
  {Starkman}}]{Zlosnik07}
{Zlosnik}, T.~G., {Ferreira}, P.~G., \& {Starkman}, G.~D. 2007, \prd, 75,
  044017

\end{thebibliography}

\label{lastpage}

\end{document}